\documentclass[conference,a4paper]{APSIPA2026}
\usepackage{amsmath}
\usepackage{graphicx}
\usepackage{subcaption}
\usepackage{multirow}
\usepackage{threeparttable}
\usepackage[backend=biber,style=ieee,]{biblatex}
\usepackage{graphicx}
\usepackage{amsmath}
\usepackage{amssymb}
\usepackage{amsfonts}
\usepackage{amsxtra}
\usepackage{mathrsfs}
\usepackage{latexsym}
\usepackage{tasks}
\usepackage{url}
\usepackage{algorithm}
\usepackage[noend]{algpseudocode}
\usepackage{xcolor}
\usepackage{booktabs}
\usepackage{tabularx}
\usepackage{arydshln}
\usepackage{listings}
\usepackage{multirow}
\usepackage{tikz}
\usepackage{pgfplots}
\usetikzlibrary{arrows.meta,calc,decorations.markings,math,shapes.geometric,positioning}

\usepackage{mathtools}
\usepackage{cleveref}

\makeatletter
\let\MYcaption\@makecaption
\makeatother
\usepackage{subcaption}
\makeatletter
\let\@makecaption\MYcaption
\makeatother
\renewcommand{\vec}[1]{\boldsymbol{#1}}
\newcommand{\mat}[1]{\boldsymbol{\mathrm{#1}}}
\newcommand{\set}[1]{\mathbb{#1}}

\newcounter{num}

\DeclareMathOperator*{\argmin}{arg~min}

\definecolor{myorange}{rgb}{0.96, 0.67, 0}
\definecolor{mygreen}{rgb}{0.01, 0.69, 0.48}

\crefname{equation}{}{}% {環境名}{単数形}{複数形} \crefで引くときの表示
\crefname{figure}{Fig.}{Figs.}% {環境名}{単数形}{複数形} \crefで引くときの表示
\crefname{table}{Table}{Tables}% {環境名}{単数形}{複数形} \crefで引くときの表示
\crefname{algorithm}{Algorithm}{Algorithm}

\crefname{equation}{}{}% {環境名}{単数形}{複数形} \crefで引くときの表示
\crefname{figure}{Fig.}{Figs.}% {環境名}{単数形}{複数形} \crefで引くときの表示
\crefname{table}{Table}{Tables}% {環境名}{単数形}{複数形} \crefで引くときの表示
\crefname{algorithm}{Algorithm}{Algorithm}
\usepackage{placeins}

\usepackage{geometry}
\usepackage{fancyhdr}

\fancypagestyle{firststyle}{
  \fancyhf{}
  \fancyhead[C]{
    2026 Asia Pacific Signal and Information Processing Association Annual Summit and
    Conference (APSIPA ASC)
  }
}

\newcommand{\addieeecopyrightnotice}{%
  \AddToShipoutPictureFG*{%
    \AtPageLowerLeft{%
      \raisebox{6mm}{%
        \makebox[\paperwidth][c]{%
          \parbox{0.94\paperwidth}{%
            \centering
            \fontsize{6}{7}\selectfont
            \textcopyright~2026 IEEE. Personal use of this material is permitted.
            Permission from IEEE must be obtained for all other uses, in any current
            or future media, including reprinting/republishing this material for
            advertising or promotional purposes, creating new collective works, for
            resale or redistribution to servers or lists, or reuse of any copyrighted
            component of this work in other works.%
          }%
        }%
      }%
    }%
  }%
}

\newcommand{\loss}{\mathcal{L}}
\newcommand{\proj}{\Pi}

\begin{document}

\addieeecopyrightnotice

\title{
Boundary-Continuous Cross-Camera RGB Mapping\\
via Hue-Split Model Trees
}

\author{
\authorblockN{
Yuma Kinoshita\authorrefmark{1} and
Hitoshi Kiya\authorrefmark{2}
}

\authorblockA{
\authorrefmark{1}
Tokai University, Japan
}

\authorblockA{
\authorrefmark{2}
Tokyo Metropolitan University, Japan
}
}

\maketitle
\thispagestyle{firststyle}
\pagestyle{empty}

\begin{abstract}
We propose a hue-split model-tree method for boundary-continuous cross-camera RGB mapping.
Cross-camera RGB mapping aims to produce consistent color representations across cameras
whose recorded RGB values differ due to sensor spectral sensitivities and image-signal processing pipelines.
A common chart-based remedy is to estimate a single global affine color correction matrix (CCM),
but such a global model cannot capture hue-specific discrepancies between cameras.
To capture that behavior, we recursively partitions
the source-camera color space along a scalar hue coordinate
and builds an model tree that stores an affine CCM at every node.
For fitting the node CCMs, we utilize a log-domain error objective.
To prevent false contours that arise from hard hue splits,
we further introduce a boundary-continuous formulation
in which the prediction is obtained
by blending the log-domain outputs of all node CCMs along the root-to-leaf path.
The path-wise blending weights are optimized under a simplex constraint
using both a chart-pair fitting loss and an explicit continuity regularizer
defined on deterministic boundary prototype pairs placed just on
either side of each learned hue threshold.
We conducted an experiment on a Canon EOS-1Ds Mark II to Canon EOS 20D mapping using
the Middlebury Registered Color Checker dataset.
The results show that
hue splitting substantially reduces log-RMSE over a single global affine CCM
and that the proposed path blending with boundary prototype regularization
simultaneously improves accuracy and suppresses chromaticity gaps
at the learned hue thresholds across two illuminants and multiple exposure conditions.
\end{abstract}


\section{Introduction}
Color is one of the most fundamental cues exploited in computer vision.
Object recognition, image segmentation, visual tracking, and scene understanding
all rely on the assumption that the color of a surface is a stable,
reproducible quantity.
In practice, however, the RGB values recorded for a given surface depend on
sensor spectral sensitivities, the optics, and the subsequent image-signal
processing pipeline \cite{ramanath2005pipeline}.
Because these factors differ from device to device,
two cameras photographing the same scene under the same illumination
can produce noticeably different RGB triplets,
which complicates multi-camera acquisition, content reuse across
devices, and reproducible image analysis.
This paper focuses on the cross-camera RGB mapping
for obtaining consistent color representations across different cameras.

A practical remedy is chart-based calibration,
where paired color measurements
from a source camera and a target camera
are used to estimate a direct RGB-to-RGB mapping.
The mapping is often implemented as a linear or affine transform
by a single global color correction matrix (CCM).
Such simple models remain attractive and widely used in calibration workflows
\cite{finlayson1997constrained}
because they are easy to estimate and
computationally light.
However, a single global CCM cannot capture
hue-specific discrepancies between cameras,
so the correction error tends to be
unevenly distributed across hues.

This paper addresses that limitation with a \emph{hue-split model tree}.
A model tree is a decision tree whose leaves hold
local regression models instead of constant predictions
\cite{quinlan1992continuous, wang1997modeltree}.
In our formulation,
every node of the tree stores a $3{\times}4$ affine CCM
that maps source-camera RGB to target-camera RGB.
We construct the tree by recursively splitting
the source-camera RGB color space along a hue coordinate:
at current leaf node,
we search the hue threshold that minimizes
the total post-split fitting loss
for adding child nodes, and
then estimate a dedicated affine CCM for each child node.
Splitting continues until the maximum log-domain error
of chart-color pairs corresponding to each leaf
falls below a tolerance.
Because different hue partitions are now served
by their own affine maps,
the tree captures hue-dependent inter-camera discrepancies
that a single global CCM cannot model,
while remaining interpretable and lightweight.

To reduce false contours due to hard hue splits,
we also propose a boundary-continuous formulation with two key ideas.
First, inference blends the log-domain estimates
from all node CCMs on the selected root-to-leaf path,
rather than using only the terminal leaf estimate.
Second, the blending weights are optimized
not only for prediction accuracy on chart pairs
but also for continuity across every hue split.
For that purpose,
we introduce deterministic boundary prototype pairs,
i.e., synthetic source-camera colors placed just to the left
and right of a split.
Penalizing their output discrepancy yields a direct,
split-aware continuity regularizer.

Our contributions are threefold:
\begin{itemize}
\item We formulate cross-camera RGB mapping as a deterministic hue-split model tree
with node-wise affine regressors in the log domain.
\item We introduce deterministic boundary prototype pairs and a path-weight
optimization scheme that explicitly penalizes discontinuities at learned hue
thresholds.
\item We validate the proposed method on the Middlebury Registered Color Checker
dataset, and it is shown that hue splitting with boundary prototype regularization
improves cross-camera mapping accuracy and boundary continuity
under multiple illuminants and exposure conditions.
\end{itemize}

\section{Related Work}
\subsection{Global Chart-Based Camera Mapping}
Classical camera characterization often maps camera RGB values
to a device-independent color space or another device
using a global regression model estimated from chart measurements.
Constrained least-squares regression and related 3$\times$3 transforms
remain a standard baseline because of their simplicity and robust behavior under
calibration constraints \cite{finlayson1997constrained}.
Higher-order polynomial mappings can reduce fitting error
by expanding the feature set beyond the raw RGB triplet \cite{hong2001polynomial}.
Comparative studies have also shown that neural networks
and polynomial transforms can offer similar accuracy,
with polynomial models often preferred for ease of training and deployment \cite{cheung2004comparative}.

More recently, root-polynomial regression has been proposed as a nonlinear extension
that retains desirable scaling behavior better than ordinary polynomial expansion
\cite{finlayson2015root}.
These methods are effective when one smooth global transform is adequate,
but they still optimize a single set of coefficients for all colors.
As a result, they do not explicitly model region-specific behavior that may arise when
inter-camera discrepancies vary with hue.

\subsection{Hue-Aware and Localized Color Correction}
Several studies have observed that partitioning the color domain can improve
approximation accuracy.
Find Andersen and Hardeberg proposed a hue-plane-preserving
approach that uses multiple 3$\times$3 matrices for different subsets of camera
responses \cite{andersen2005hue}.
More recently, Li \emph{et al.} showed that grouping training samples by hue angle
can improve conversion accuracy for wide-color-gamut cameras \cite{li2023groups}.
These results motivate local modeling: if different hue
regions exhibit different mapping behavior,
a set of local regressors can be more accurate
than a single global matrix.

However, local color correction also introduces a practical problem.
When neighboring regions are calibrated independently,
their predictions need not agree near the partition boundary.
For image-wide application,
even small output jumps can produce visible false contours
on smooth gradients.
Therefore, a localized mapping should
ideally improve fidelity without sacrificing continuity.

\subsection{Model Trees and Smooth Tree-Based Prediction}
Model trees address nonlinear regression by recursively partitioning the input space and
assigning a simple regressor to each region.
Quinlan's M5 framework and its later formulation
by Wang and Witten established model trees as an interpretable alternative
to purely global regression \cite{quinlan1992continuous, wang1997modeltree}.
Because each leaf contains a local parametric model,
model trees are well suited to problems
that are globally nonlinear but locally simple.

A related line of work seeks smoother behavior
by softly combining local experts.
Hierarchical mixtures of experts replace hard decisions
with probabilistic gating and weighted expert outputs \cite{jordan1994hme}.
Such models provide smooth transitions,
but they generally require learning both gating and expert functions jointly
and are not tailored to deterministic hue partitions derived from chart calibration.

Our approach combines the interpretability of a deterministic hue-split tree
with a lightweight smoothing mechanism.
Instead of replacing the tree with a fully
probabilistic gating network,
we keep the explicit hue intervals and optimize path-wise blending weights
after the local affine models have been estimated.
Moreover, continuity is enforced by boundary prototype pairs
that directly target the
split locations responsible for visible discontinuities.

\section{Proposed Method}
Given $K$ paired chart measurements from a source camera
and a target camera,
\begin{equation}
  \mathcal{D} = \{(\vec{x}_i, \vec{y}_i)\}_{i=1}^{K},
  \qquad \vec{x}_i, \vec{y}_i \in \set{R}_{+}^{3},
  \label{eq:data}
\end{equation}
our goal is to estimate a mapping
$f\colon \set{R}_{+}^{3}\to\set{R}_{+}^{3}$ that satisfies
two requirements simultaneously:
\begin{enumerate}
  \item \textbf{Accuracy.}
        $f(\vec{x}_i)$ reproduces the target-camera RGB $\vec{y}_i$
        for every chart pair.
  \item \textbf{Boundary continuity.}
        $f$ varies smoothly across the entire source-camera color domain.
\end{enumerate}

To pursue both requirements,
we model $f$ as a piecewise affine transform
organized by a hue-split model tree.
We augment each input with a bias term,
$\tilde{\vec{x}}_i
  = [\vec{x}_i^{\top}\ 1]^{\top}\!\in\set{R}^{4}$,
and assign an affine matrix
$\tilde{\mat{M}}\in\set{R}^{3\times 4}$
to every tree node so that
a node predicts $\tilde{\mat{M}}\tilde{\vec{x}}$.
We also define a scalar hue coordinate $h(\vec{x})$
extracted from the source-camera RGB
after white balancing and HSV conversion;
$h$ serves as the sole splitting variable of the tree.
\begin{figure}[t]
\centering
\includegraphics[width=\linewidth]{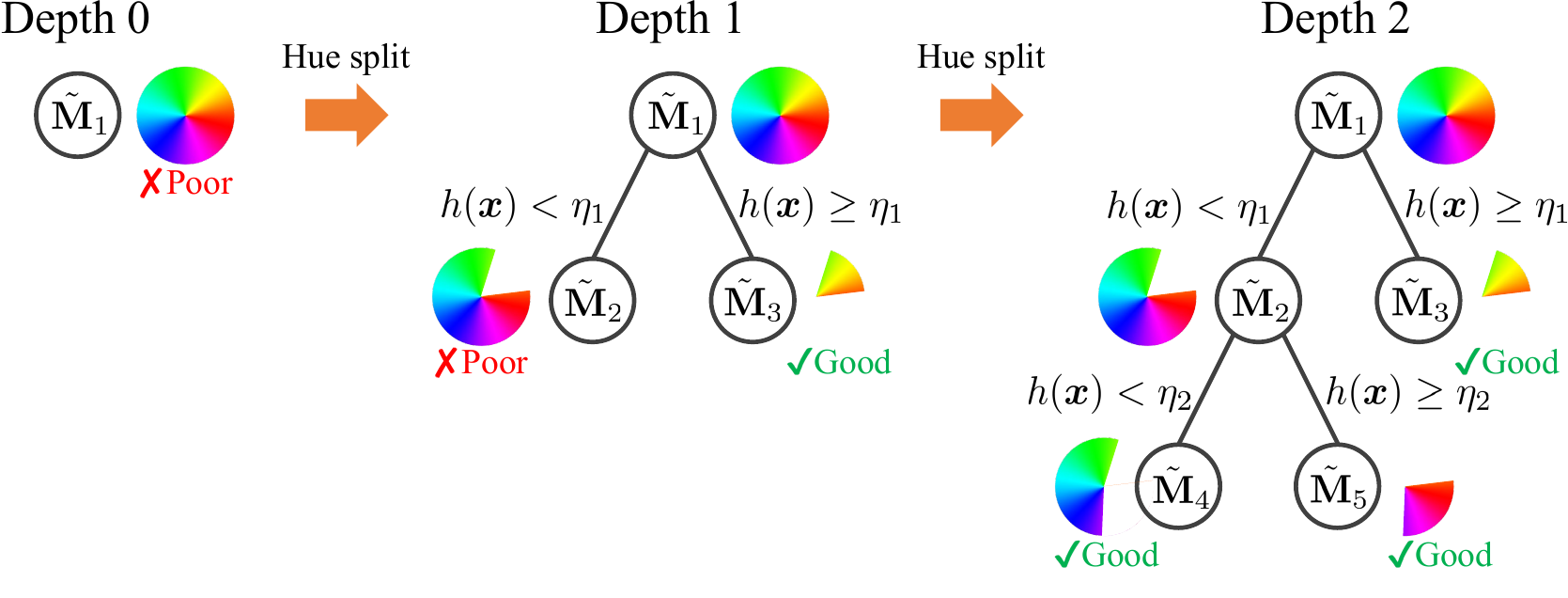}
\caption{Building hue-split model tree}
\label{fig:building_tree}
\end{figure}
\begin{figure}[t]
\centering
\includegraphics[width=0.6\linewidth]{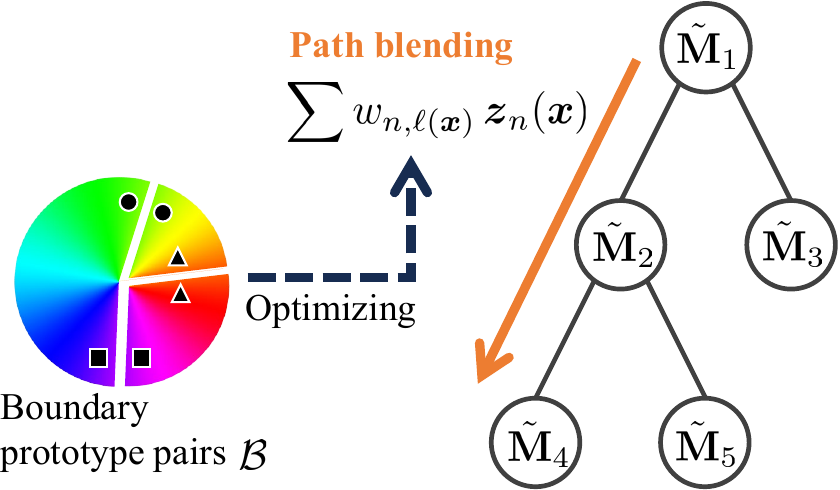}
\caption{Path blending with boundary prototype regularization}
\label{fig:path_blending}
\end{figure}

\subsection{Hue Coordinate}
\label{ssec:hue_coord}
The tree partitions the source-camera color space
by a scalar hue coordinate $h(\vec{x})$.
We obtain $h$ from the source RGB $\vec{x}$
via white balancing followed
by conversion to HSV color space.
White balancing is simply done using
the source RGB $\vec{x}_\mathrm{white}$ of a white patch in the chart as
\begin{equation}
\vec{x}^{\mathrm{wb}}
  = \vec{x} \oslash \vec{x}_\mathrm{white},
\label{eq:white_balance}
\end{equation}
where $\oslash$ denotes element-wise division.

Writing $\vec{x}^{\mathrm{wb}} = (x_R,x_G,x_B)$ and defining
\begin{align}
  C_{\max} &= \max(x_R,x_G,x_B), \\
  C_{\min} &= \min(x_R,x_G,x_B), \\
  \Delta &= C_{\max} - C_{\min},
\end{align}
conversion to the value $V$, saturation $S$, and hue $H$ are defined as
\begin{align}
V &= C_{\max}, \label{eq:hsv_v} \\
S &= \begin{cases}
  \Delta / C_{\max} & \text{if } C_{\max} > 0, \\
  0 & \text{otherwise,}
\end{cases} \label{eq:hsv_s} \\
H &= \begin{cases}
  0^{\circ}
    & \text{if } \Delta = 0, \\[2pt]
  60^{\circ}\!\times\!\dfrac{x_G-x_B}{\Delta} \bmod 360^{\circ}
    & \text{if } C_{\max}=x_R, \\[4pt]
  60^{\circ}\!\times\!\dfrac{x_B-x_R}{\Delta}+120^{\circ}
    & \text{if } C_{\max}=x_G, \\[4pt]
  60^{\circ}\!\times\!\dfrac{x_R-x_G}{\Delta}+240^{\circ}
    & \text{if } C_{\max}=x_B.
\end{cases} \label{eq:hsv_h}
\end{align}

We define the hue coordinate as
\begin{equation}
h(\vec{x}) =
\begin{cases}
  H & \text{if } S \geq s_{\min}, \\
  0 & \text{otherwise.}
\end{cases}
\end{equation}
The hue coordinate $h(x)$ is used as a device-dependent feature for branch selection
and is not intended to represent a colorimetrically accurate estimate of perceptual hue.
Although $S$ and $V$ are not used for tree splitting,
they are used to construct the boundary prototype pairs
in Section~\ref{ssec:boundary}.

\subsection{Hue-Split Model-Tree Construction}
For a node $n$ with chart-color index set
$\mathcal{S}_{n} \subset \{1,\ldots,K\}$,
we first fit its affine CCM $\tilde{\mat{M}}_{n}$
by following the log-domain error minimization:
\begin{align}
  \tilde{\mat{M}}_{n} &=
  \tilde{\mat{M}}^{\star}(\mathcal{S}_n)
  = \argmin_{\tilde{\mat{M}}}\;\loss(\tilde{\mat{M}};\,\mathcal{S}),
  \label{eq:fit_node} \\
  \loss(\tilde{\mat{M}};\,\mathcal{S})
  &= \frac{1}{3|\mathcal{S}|}
    \sum_{i\in\mathcal{S}}
    \bigl\|
      \log_{2}\!\bigl([\tilde{\mat{M}}\tilde{\vec{x}}_i]_{+}+\epsilon\bigr)
      - \log_{2}(\vec{y}_i+\epsilon)
    \bigr\|_2^2,
  \label{eq:node_loss}
\end{align}
where $[\cdot]_{+}$ denotes the element-wise positive-part operator
(i.e., $[z]_{+} = \max(z, 0)$ for a scalar $z$)
and $\epsilon>0$ is a small constant to avoid zero arguments to the logarithm.
In our implementation, the affine CCM is optimized with a Gauss--Newton method
initialized from the least-squares solution in the linear domain.

We then evaluate the node's maximum training error,
\begin{equation}
\varepsilon_{\max}(n) = \max_{i \in \mathcal{S}_{n}}
\frac{1}{3}
\left\| \log_{2}\!\left([\tilde{\mat{M}}_{n}\tilde{\vec{x}}_i]_{+}+\epsilon\right)
- \log_{2}(\vec{y}_i+\epsilon) \right\|_2^2.
\label{eq:max_error}
\end{equation}
If $\varepsilon_{\max}(n) \leq \tau$,
or if the node violates practical stopping
conditions such as a minimum leaf size or a maximum depth,
the node becomes a leaf.

Otherwise, the node is split by a hue threshold $\eta$.
Let
\begin{align}
\mathcal{S}_{L}(\eta) &= \{ i \in \mathcal{S}_{n} \mid h(\vec{x}_i) < \eta \}, \\
\mathcal{S}_{R}(\eta) &= \{ i \in \mathcal{S}_{n} \mid h(\vec{x}_i) \geq \eta \},
\end{align}
under an assumption that $\{\vec{x}_i\}$ are sorted by hue.
Among admissible candidate thresholds $\eta \in \mathcal{T}_{n}$, formed by the
midpoints of sorted hue values, we choose the one that minimizes the total
post-split loss,
\begin{equation}
\begin{aligned}
\eta_{n} = {} & \argmin_{\eta \in \mathcal{T}_{n}}
\Big(
|\mathcal{S}_{L}(\eta)|\,\loss\big(
\tilde{\mat{M}}^{\star}(\mathcal{S}_{L}(\eta)); \mathcal{S}_{L}(\eta)\big)
\\
& \quad + |\mathcal{S}_{R}(\eta)|\,\loss\big(
\tilde{\mat{M}}^{\star}(\mathcal{S}_{R}(\eta)); \mathcal{S}_{R}(\eta)\big)
\Big).
\end{aligned}
\label{eq:split_search}
\end{equation}

The above procedure is started at the root node
with $\mathcal{S}_{1} = \{1,\ldots,K\}$
and it is then applied recursively to the left and right children.
This produces a deterministic hue-split model tree in which each node stores
its own affine CCM.

Compared with a single CCM, the tree can approximate different local mappings in
different hue intervals.
Compared with arbitrary nonlinear regressors, the resulting
representation remains interpretable: the tree explicitly reveals
where the source-camera hue domain has been partitioned
and which local affine map is responsible for each region.

\subsection{Path Blending in the Log Domain}
Using only the leaf CCM can make the output discontinuous at a split boundary
because adjacent leaves are fitted almost independently.
We therefore blend all estimates on the root-to-leaf path.
Let $\ell(\vec{x})$ be the leaf selected by the
hue-based routing rule for input $\vec{x}$,
and let $\mathcal{P}_{\ell(\vec{x})}$
denote the set of nodes on the corresponding path.

For each node $n$ on that path,
we define the local log-domain prediction
\begin{equation}
\vec{z}_{n}(\vec{x}) = \log_{2}\!\left([\tilde{\mat{M}}_{n}\tilde{\vec{x}}]_{+}+\epsilon\right),
\label{eq:local_log_pred}
\end{equation}
and we blend those predictions with path-specific nonnegative weights
$w_{n, \ell(\vec{x})} \geq 0$
as the final log-domain estimate $\hat{\vec{z}}(\vec{x})$,
\begin{equation}
\bar{\vec{z}}(\vec{x}) = \sum_{n \in \mathcal{P}_{\ell(\vec{x})}} w_{n, \ell(\vec{x})}\,\vec{z}_{n}(\vec{x})
\quad \mathrm{s.t.} \ 
\sum_{n \in \mathcal{P}_{\ell(\vec{x})}} w_{n, \ell(\vec{x})} = 1.
\label{eq:path_blend}
\end{equation}
The RGB prediction in the target-camera domain is then obtained by inverse log mapping,
\begin{equation}
\bar{\vec{y}}(\vec{x}) = 2^{\bar{\vec{z}}(\vec{x})} - \epsilon,
\label{eq:inverse_log}
\end{equation}
followed by clipping to the valid target RGB range.

\subsection{Blending Weight Optimization with Continuity Regularization}
\label{ssec:boundary}
To optimize the weights $\{w_n\}$,
we use not only the chart-fitting error
but also an explicit boundary continuity term.
Consider an internal node with hue threshold $\eta$.
For that threshold,
we construct a deterministic set of boundary prototype pairs,
\begin{equation}
\mathcal{B}_{\eta} = \{(\vec{b}^{-}_{\eta,m}, \vec{b}^{+}_{\eta,m})\}_{m=1}^{M_{\eta}},
\end{equation}
where each pair is generated from a source color
on the split boundary $\eta$
by a small hue perturbation $\delta$ in either direction:
one is perturbed to $\eta-\delta$ and the other to
$\eta+\delta$, with saturation and value held fixed.
Thus, the pair straddles the split boundary
while remaining otherwise comparable.

The objective for weight optimization consists of three terms:
a fitting term that encourages accurate chart prediction,
a regularizer that penalizes discontinuity at the hue boundaries,
and a small $\ell_2$ stabilization term.
The total objective becomes
\begin{equation}
J(\vec{w}) = \frac{1}{2}\mathcal{E}(\vec{w})
+ \frac{\lambda}{2}\mathcal{R}(\vec{w})
+ \frac{\xi}{2}\|\vec{w}\|_2^2,
\label{eq:total_obj}
\end{equation}
where $\lambda > 0$ controls the trade-off between prediction accuracy and continuity,
and $\xi > 0$ is a small stabilization parameter.
The fitting term $\mathcal{E}$ is given by the average log-domain MSE over the chart pairs as
\begin{equation}
\mathcal{E}(\vec{w}) = \frac{1}{3K}
\sum_{i=1}^{K}
\left\| \bar{\vec{z}}(\vec{x}_i) - \log_{2}(\vec{y}_i+\epsilon) \right\|_2^2,
\label{eq:data_term}
\end{equation}
while the boundary continuity term $\mathcal{R}$ is given by the average squared log-domain jump
over the boundary prototype pairs as
\begin{equation}
\mathcal{R}(\vec{w}) = \frac{1}{3|\mathcal{B}|}
\sum_{(\vec{b}^{-},\vec{b}^{+}) \in \mathcal{B}}
\left\| \bar{\vec{z}}(\vec{b}^{-}) - \bar{\vec{z}}(\vec{b}^{+}) \right\|_2^2,
\label{eq:cont_term}
\end{equation}
where $\mathcal{B} = \bigcup_{\eta} \mathcal{B}_{\eta}$ is
the union of all boundary prototype pairs across all splits.

For a fixed tree and fixed node CCMs,
$J(\vec{w})$ is convex in the weights
because both \eqref{eq:data_term} and \eqref{eq:cont_term} are quadratic in $\vec{w}$
and the feasible set induced by the simplex constraint in Eq.~\eqref{eq:path_blend}
is convex.
Therefore, we optimize
$\vec{w}$ by projected gradient descent as
\begin{equation}
\vec{w}^{(t+1)}
= \proj_{\mathcal{C}}\!\left( \vec{w}^{(t)} - \alpha\, \nabla J(\vec{w}^{(t)}) \right),
\label{eq:pgd}
\end{equation}
where $\proj_{\mathcal{C}}$ is the projection onto
the set $\mathcal{C} = \{\vec{w} \mid w_n \geq 0,\,
\sum_{n \in \mathcal{P}_{\ell}} w_n = 1,\, \forall \ell\}$
and $\alpha > 0$ is a step size.

\section{Experiment}
We conducted an experiment using real data to answer the following questions:
1) whether hue-wise local modeling improves cross-camera RGB mapping accuracy
over a single global affine CCM,
and 2) whether the proposed path blending with boundary prototype regularization
reduces discontinuities at learned hue thresholds.

\subsection{Experimental Conditions}
We evaluated accuracy and boundary continuity of color mapping
from Canon EOS-1Ds Mark II to Canon EOS 20D,
using a raw-image pair from the Middlebury Registered Color Checker dataset
\cite{middleburycolor}.
In the dataset, we used the \texttt{checker140s-RAW-JPG} archive,
kept the RAW-derived linear PNGs with \texttt{wb1},
extracted the inner 96 Digital ColorChecker SG patches.
We built our model tree using RGB values of the 96 patches at the exposure condition of 0 EV 
while testing RGB values at $v \in \{-1,0,1\}$ EV
under both available light conditions: \texttt{i1} and \texttt{i2}.

We compared the following three mapping methods:
\begin{enumerate}
  \item Single global affine CCM (NoSplit-NoBlend):
    using $\tilde{\mat{M}}_{1}$ fitted at the root node
    for all input colors without hue splitting.
  \item Hard hue-split model tree without path blending (HueSplit-NoBlend):
    using only the leaf CCM for prediction
    without any path blending or continuity regularization.
  \item Hue-split model tree with fixed M5-style path blending (HueSplit-M5Blend):
    using the M5 heuristic~\cite{quinlan1992continuous} for path blending weights
    without any continuity regularization.
  \item Proposed hue-split model tree with path blending (HueSplit-OptimizedBlend):
    using the path blending and boundary prototype regularization
    with $\lambda \in \{0, 0.1, 1\}$.
\end{enumerate}
All tree-based methods shared the same tree structure
including hue threshold and node CCMs.
During building the tree,
we set the maximum depth to 2, the minimum leaf size to 4,
and the max-error tolerance $\tau$ to 0 (i.e., no early stopping based on error).

For learned path blending,
we use $\epsilon=10^{-6}$, $\xi=10^{-6}$,
and a set $\mathcal{B}$ of deterministic boundary prototype pairs
generated with hue perturbation $\delta=1^{\circ}$
on a $5\times4$ saturation-value grid,
where $S\in[0.05,1.0]$ and $V\in[0.04,0.16]$.
The achromatic threshold is set to $s_{\min}=10^{-8}$.

Accuracy is measured by the patch-wise log-root-mean-squared error (log-RMSE) as
\begin{equation}
d(\hat{\vec{y}},\vec{y})=
\sqrt{\frac{1}{3}}\left\|\log_{2}(\hat{\vec{y}}+\epsilon)-\log_{2}(\vec{y}+\epsilon)\right\|_{2},
\label{eq:metric_logrmse}
\end{equation}
and boundary smoothness is evaluated by the average boundary jump
\begin{equation}
B = \frac{1}{|\mathcal{B}|}\sum_{(\vec{b}^{-},\vec{b}^{+})\in\mathcal{B}}
\sqrt{\frac{1}{3}}\left\|\hat{\vec{z}}(\vec{b}^{-})-\hat{\vec{z}}(\vec{b}^{+})\right\|_{2},
\label{eq:metric_boundary}
\end{equation}
where $\mathcal{B}$ is the same set of deterministic boundary prototype pairs
used for regularization in Eq.~\eqref{eq:cont_term}.

\subsection{Results}
\begin{figure}[t]
\centering
\begin{subfigure}[t]{0.32\columnwidth}
\centering
\includegraphics[width=\linewidth]{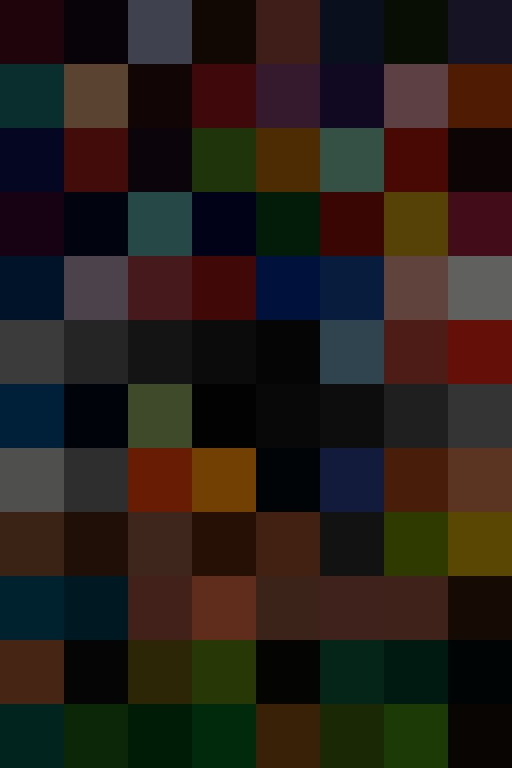}
\caption{\texttt{Input chart}}
\end{subfigure}
\hfill
\begin{subfigure}[t]{0.32\columnwidth}
\centering
\includegraphics[width=\linewidth]{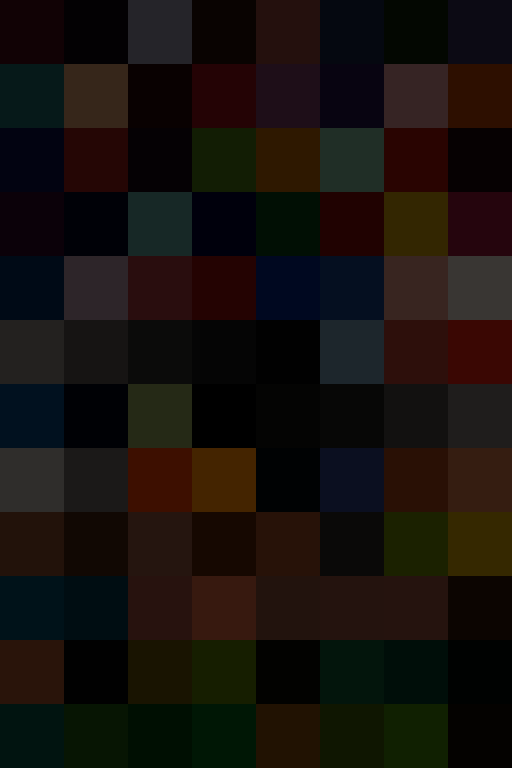}
\caption{\texttt{Target chart}}
\end{subfigure}
\hfill
\begin{subfigure}[t]{0.32\columnwidth}
\centering
\includegraphics[width=\linewidth]{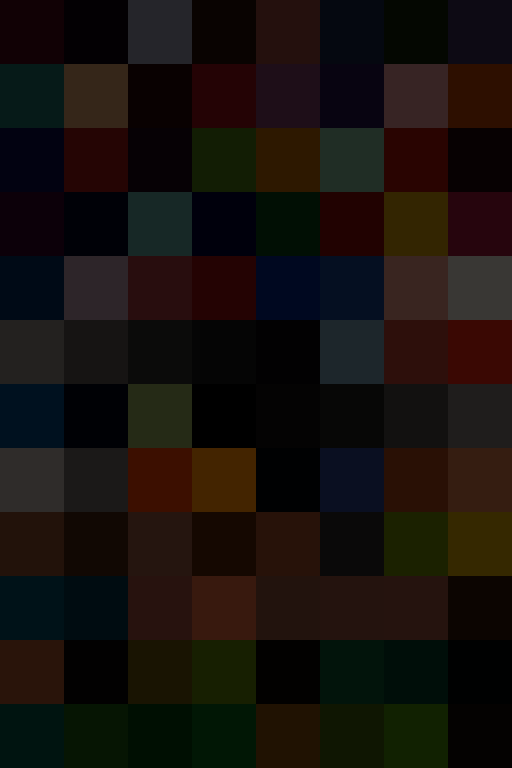}
\caption{\texttt{NoSplit-} \texttt{NoBlend}}
\end{subfigure}
\par\medskip
\begin{subfigure}[t]{0.32\columnwidth}
\centering
\includegraphics[width=\linewidth]{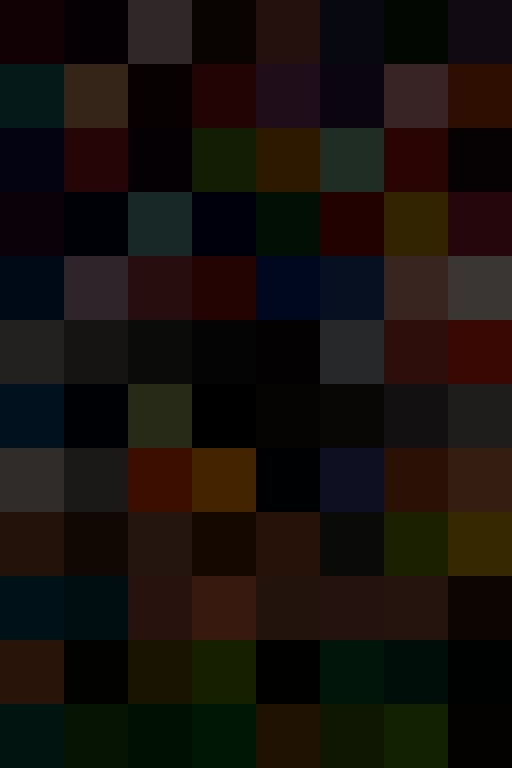}
\caption{\texttt{HueSplit-} \texttt{NoBlend}}
\end{subfigure}
\hfill
\begin{subfigure}[t]{0.32\columnwidth}
\centering
\includegraphics[width=\linewidth]{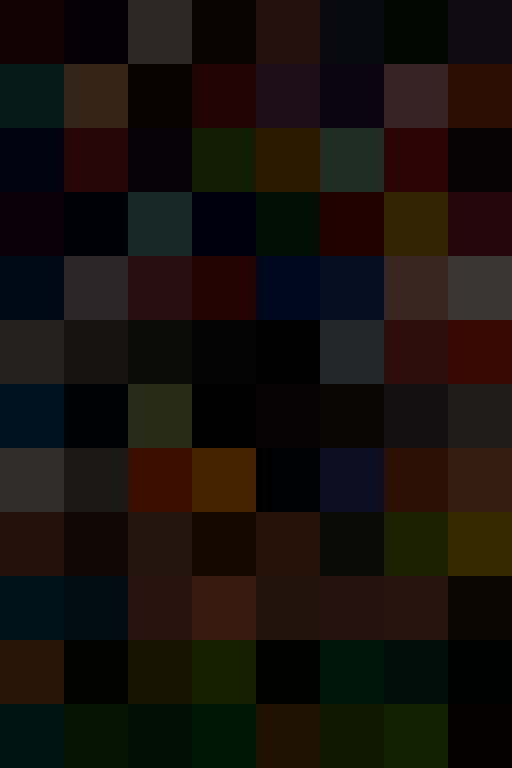}
\caption{\texttt{HueSplit-} \texttt{M5Blend}~\cite{quinlan1992continuous}}
\end{subfigure}
\hfill
\begin{subfigure}[t]{0.32\columnwidth}
\centering
\includegraphics[width=\linewidth]{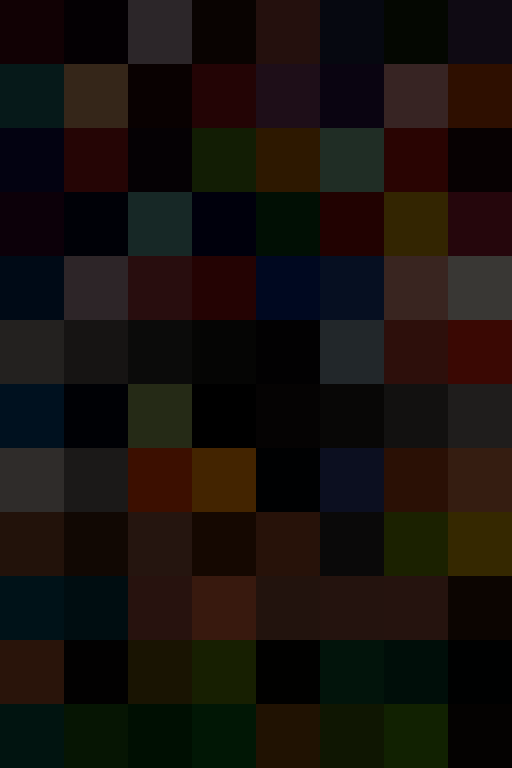}
\caption{\texttt{HueSplit-} \texttt{OptimizedBlend} ($\lambda=1.0$)}
\end{subfigure}
\caption{
Qualitative comparison of corrected chart patches for the Middlebury Color Checker dataset
under light \texttt{i1} at 0 EV and depth 2.
}
\label{fig:middlebury_chart}
\end{figure}

\begin{figure}[t]
\centering
\begin{subfigure}[t]{0.32\columnwidth}
\centering
\includegraphics[width=\linewidth]{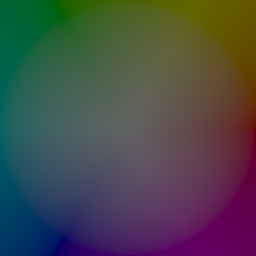}
\caption{\texttt{Input gradient} (+2 EV)}
\end{subfigure}
\hfill
\begin{subfigure}[t]{0.32\columnwidth}
\centering
\includegraphics[width=\linewidth]{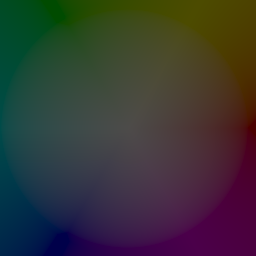}
\caption{\texttt{NoSplit-} \texttt{NoBlend}}
\end{subfigure}
\hfill
\begin{subfigure}[t]{0.32\columnwidth}
\centering
\includegraphics[width=\linewidth]{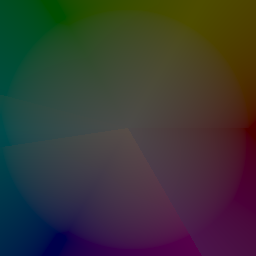}
\caption{\texttt{HueSplit-} \texttt{NoBlend}}
\end{subfigure}
\par\medskip
\begin{subfigure}[t]{0.32\columnwidth}
\centering
\includegraphics[width=\linewidth]{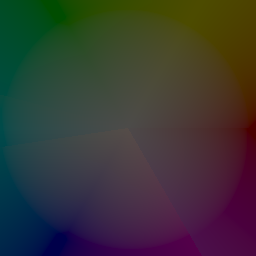}
\caption{\texttt{HueSplit-} \texttt{M5Blend}}
\end{subfigure}
\hfill
\begin{subfigure}[t]{0.32\columnwidth}
\centering
\includegraphics[width=\linewidth]{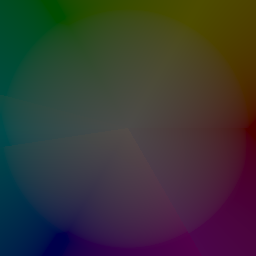}
\caption{\texttt{HueSplit-} \texttt{OptimizedBlend} ($\lambda=0.1$)}
\end{subfigure}
\hfill
\begin{subfigure}[t]{0.32\columnwidth}
\centering
\includegraphics[width=\linewidth]{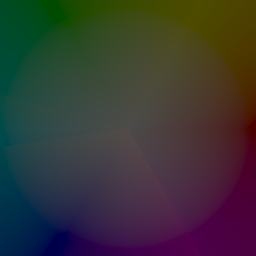}
\caption{\texttt{HueSplit-} \texttt{OptimizedBlend} ($\lambda=1.0$)}
\end{subfigure}
\caption{
Chromaticity-gap visualization for methods trained on the Middlebury Color Checker dataset under light
\texttt{i1}, by applying the methods to a +2 EV gradient image.
The depth is 2 for all tree-based methods.
}
\label{fig:middlebury_gap}
\end{figure}

Figures \ref{fig:middlebury_chart} and \ref{fig:middlebury_gap}
visualize the corresponding 0 EV correction under \texttt{i1}
and the +2 EV chromaticity gap under the same light at depth 2, respectively.
From \cref{fig:middlebury_gap},
we can see that \texttt{HueSplit-NoBlend} created a visible chromaticity gap at the learned hue threshold,
and M5 and optimized blends with $\lambda=0.1$ and $\lambda=1.0$ reduced the gap.

Table~\ref{tab:middlebury_i1} summarizes the accuracy and boundary continuity under light
condition \texttt{i1}.
At the training exposure of 0 EV,
hue splitting yields a substantial reduction in log-RMSE over the single global affine CCM:
the error drops from 0.8426 for \texttt{NoSplit-NoBlend} to 0.4977 at depth 1
and further to 0.2110 at depth 2 for \texttt{HueSplit-NoBlend}.
This improvement shows that
hue-wise local modeling captures non-uniform color shifts
that a single global affine map cannot represent.
Comparing the three variants of \texttt{HueSplit-OptimizedBlend} with \texttt{HueSplit-NoBlend}
under the same depth condition,
the proposed HueSplit-OptimizedBlend consistently improves both accuracy and continuity
while \texttt{HueSplit-M5Blend} sometimes degrades accuracy relative to \texttt{HueSplit-NoBlend}.
For this results,
we can see that the boundary prototype regularization effectively
suppresses discontinuities at the learned hue thresholds without sacrificing accuracy.
\begin{table*}[t]
\centering
\caption{
Log-RMSE and boundary jump $B$ of cross-camera RGB mapping
from Canon EOS-1Ds Mark II to EOS 20D
on the Middlebury Color Checker dataset under illuminant \texttt{i1}.
Lower is better for both metrics.
Models are trained at 0 EV.
"Depth" indicates the test-time depth at which the root-to-leaf
path through the model tree is truncated.
}
\label{tab:middlebury_i1}
\small
\begin{tabular}{llcccccc}
\hline
Method & Depth & \multicolumn{2}{c}{$-1$ EV} & \multicolumn{2}{c}{$0$ EV} & \multicolumn{2}{c}{$+1$ EV} \\
 &  & log-RMSE & $B$ [EV] & log-RMSE & $B$ [EV] & log-RMSE & $B$ [EV] \\
\hline
\texttt{NoSplit-NoBlend} & -- & 2.6888 & -- & 0.8426 & -- & \textbf{1.2563} & -- \\
\texttt{HueSplit-NoBlend} & 1 & 2.5118 & 0.1555 & 0.4977 & 0.0766 & 1.2751 & 0.0590 \\
\texttt{HueSplit-M5Blend}~\cite{quinlan1992continuous} & 1 & 2.5096 & \textbf{0.1508} & 0.6389 & \textbf{0.0714} & 1.2728 & 0.0544 \\
\texttt{HueSplit-OptimizedBlend} ($\lambda=0$) & 1 & 2.5060 & 0.1814 & 0.4852 & 0.0755 & 1.2674 & \textbf{0.0543} \\
\texttt{HueSplit-OptimizedBlend} ($\lambda=0.1$) & 1 & 2.5060 & 0.1814 & 0.4852 & 0.0755 & 1.2674 & \textbf{0.0543} \\
\texttt{HueSplit-OptimizedBlend} ($\lambda=1.0$) & 1 & 2.5060 & 0.1814 & 0.4852 & 0.0755 & 1.2674 & \textbf{0.0543} \\
\texttt{HueSplit-NoBlend} & 2 & 3.2930 & 2.3387 & 0.2110 & 0.7737 & 1.3893 & 0.8382 \\
\texttt{HueSplit-M5Blend}~\cite{quinlan1992continuous} & 2 & 3.2416 & 1.9803 & 0.3588 & 0.5482 & 1.3693 & 0.4659 \\
\texttt{HueSplit-OptimizedBlend} ($\lambda=0$) & 2 & 2.6210 & 2.2097 & \textbf{0.1653} & 0.6813 & 1.3224 & 0.7451 \\
\texttt{HueSplit-OptimizedBlend} ($\lambda=0.1$) & 2 & 2.4046 & 1.8609 & 0.1687 & 0.5715 & 1.3192 & 0.6440 \\
\texttt{HueSplit-OptimizedBlend} ($\lambda=1.0$) & 2 & \textbf{2.2118} & 1.1992 & 0.1810 & 0.3753 & 1.3132 & 0.4546 \\
\hline
\end{tabular}
\end{table*}

Table~\ref{tab:middlebury_i2} reports the same comparison under
light condition \texttt{i2}, and overall the same trends are observed.
At the training exposure of 0 EV,
hue splitting again yields a substantial reduction in log-RMSE
over the single global affine CCM:
the error drops from 1.5016 for \texttt{NoSplit-NoBlend}
to 1.3138 at depth 1 and further to 1.1565 at depth 2
for \texttt{HueSplit-NoBlend}.
Comparing the three variants of \texttt{HueSplit-OptimizedBlend}
with \texttt{HueSplit-NoBlend} under the same depth condition,
the proposed method consistently improves both accuracy and continuity
at depth 2 across all exposures,
whereas \texttt{HueSplit-M5Blend} provides at best a marginal change
relative to \texttt{HueSplit-NoBlend}.
A notable difference from light \texttt{i1} appears at depth -1 EV,
where \texttt{HueSplit-NoBlend} (1.7858) is in fact worse
than the single global CCM (1.5766);
the proposed method with $\lambda=1.0$ recovers this loss to 1.6142,
and at depth 2 it achieves the best log-RMSE of 1.5553,
outperforming the global baseline.
These results confirm that
the boundary prototype regularization remains effective
under a different illumination.
\begin{table*}[t]
\centering
\caption{
Log-RMSE and boundary jump $B$ of cross-camera RGB mapping
from Canon EOS-1Ds Mark II to EOS 20D
on the Middlebury Color Checker dataset under illuminant \texttt{i2}.
Lower is better for both metrics.
Models are trained at 0 EV.
"Depth" indicates the test-time depth at which the root-to-leaf
path through the model tree is truncated.
}
\label{tab:middlebury_i2}
\small
\begin{tabular}{llcccccc}
\hline
Method & Depth & \multicolumn{2}{c}{$-1$ EV} & \multicolumn{2}{c}{$0$ EV} & \multicolumn{2}{c}{$+1$ EV} \\
 &  & log-RMSE & $B$ [EV] & log-RMSE & $B$ [EV] & log-RMSE & $B$ [EV] \\
\hline
\texttt{NoSplit-NoBlend} & -- & 1.5766 & -- & 1.5016 & -- & 2.2360 & -- \\
\texttt{HueSplit-NoBlend} & 1 & 1.7858 & 0.8853 & 1.3138 & 0.4597 & 2.2355 & 0.2395 \\
\texttt{HueSplit-M5Blend}~\cite{quinlan1992continuous} & 1 & 1.6274 & \textbf{0.4100} & 1.3197 & 0.1470 & 2.2354 & 0.0833 \\
\texttt{HueSplit-OptimizedBlend} ($\lambda=0$) & 1 & 1.7858 & 0.8853 & 1.3138 & 0.4597 & 2.2355 & 0.2395 \\
\texttt{HueSplit-OptimizedBlend} ($\lambda=0.1$) & 1 & 1.6359 & 0.6327 & 1.3152 & 0.2573 & 2.2357 & 0.1406 \\
\texttt{HueSplit-OptimizedBlend} ($\lambda=1.0$) & 1 & 1.6142 & 0.5439 & 1.3165 & \textbf{0.1265} & 2.2362 & \textbf{0.0741} \\
\texttt{HueSplit-NoBlend} & 2 & 1.7731 & 1.7012 & 1.1565 & 0.9983 & 2.2095 & 0.5067 \\
\texttt{HueSplit-M5Blend}~\cite{quinlan1992continuous} & 2 & 1.7677 & 1.5259 & 1.1555 & 0.8675 & 2.2072 & 0.4253 \\
\texttt{HueSplit-OptimizedBlend} ($\lambda=0$) & 2 & 1.7484 & 1.3703 & \textbf{1.1390} & 0.8061 & 2.2082 & 0.4114 \\
\texttt{HueSplit-OptimizedBlend} ($\lambda=0.1$) & 2 & 1.5978 & 0.9998 & 1.1452 & 0.4790 & \textbf{2.1993} & 0.2677 \\
\texttt{HueSplit-OptimizedBlend} ($\lambda=1.0$) & 2 & \textbf{1.5553} & 0.9077 & 1.1496 & 0.3668 & 2.1998 & 0.2197 \\
\hline
\end{tabular}
\end{table*}

\section{Conclusion}
We presented a boundary-continuous hue-split model tree for cross-camera RGB mapping.
The tree recursively partitions the source-camera color space
along a hue coordinate and stores a node-wise $3 \times 4$ affine CCM fitted in the log domain,
while a path-blending scheme with deterministic boundary prototype pairs
explicitly penalizes log-domain jumps at learned hue thresholds.
Experiments on a Canon EOS-1Ds Mark II to Canon EOS 20D mapping
using the Middlebury Registered Color Checker dataset showed that
hue splitting substantially reduces log-RMSE over a single global affine CCM,
and that the proposed boundary prototype regularization consistently improves
both accuracy and continuity at depth 2 across all tested exposures and both illuminants,
whereas the M5-style fixed blending provided only marginal continuity gains.

Limitations of this work include the evaluation on a single camera pair with 96 chart patches
under two illuminants.
Future work will validate the method on a wider range of camera pairs and illuminants,
compare with broader nonlinear baselines,
and assess perceptual quality on full-resolution natural images.
We will also examine whether incorporating sensor spectral sensitivities
and calibrated color-balance transforms improves illuminant robustness and hue-split stability.

\printbibliography

@article{ramanath2005pipeline,
  author  = {Rajeev Ramanath and Wesley E. Snyder and Youngjun Yoo and Mark S. Drew},
  title   = {Color image processing pipeline},
  journal = {IEEE Signal Processing Magazine},
  volume  = {22},
  number  = {1},
  pages   = {34--43},
  year    = {2005},
  doi     = {10.1109/MSP.2005.1407713}
}

@article{finlayson1997constrained,
  author  = {Graham D. Finlayson and Mark S. Drew},
  title   = {Constrained least-squares regression in color spaces},
  journal = {Journal of Electronic Imaging},
  volume  = {6},
  number  = {4},
  pages   = {484--493},
  year    = {1997},
  doi     = {10.1117/12.278080}
}

@article{hong2001polynomial,
  author  = {Guowei Hong and M. Ronnier Luo and Peter A. Rhodes},
  title   = {A study of digital camera colorimetric characterization based on polynomial modeling},
  journal = {Color Research \& Application},
  volume  = {26},
  number  = {1},
  pages   = {76--84},
  year    = {2001}
}

@article{cheung2004comparative,
  author  = {T. L. V. Cheung and Stephen Westland and David R. Connah and Caterina Ripamonti},
  title   = {A comparative study of the characterisation of colour cameras by means of neural networks and polynomial transforms},
  journal = {Coloration Technology},
  volume  = {120},
  pages   = {19--25},
  year    = {2004},
  doi     = {10.1111/j.1478-4408.2004.tb00201.x}
}

@inproceedings{andersen2005hue,
  author    = {Casper Find Andersen and Jon Yngve Hardeberg},
  title     = {Colorimetric Characterization of Digital Cameras Preserving Hue Planes},
  booktitle = {Proceedings of the IS\&T/SID 13th Color Imaging Conference},
  pages     = {141--146},
  year      = {2005},
  doi       = {10.2352/CIC.2005.13.1.art00028}
}

@article{finlayson2015root,
  author  = {Graham D. Finlayson and Michal Mackiewicz and Anya Hurlbert},
  title   = {Color correction using root-polynomial regression},
  journal = {IEEE Transactions on Image Processing},
  volume  = {24},
  number  = {5},
  pages   = {1460--1470},
  year    = {2015},
  doi     = {10.1109/TIP.2015.2405336}
}

@inproceedings{quinlan1992continuous,
  author    = {Ross J. Quinlan},
  title     = {Learning with Continuous Classes},
  booktitle = {Proceedings of the 5th Australian Joint Conference on Artificial Intelligence},
  publisher = {World Scientific},
  address   = {Singapore},
  pages     = {343--348},
  year      = {1992}
}

@inproceedings{wang1997modeltree,
  author    = {Yong Wang and Ian H. Witten},
  title     = {Induction of model trees for predicting continuous classes},
  booktitle = {Poster Papers of the 9th European Conference on Machine Learning},
  publisher = {Springer},
  year      = {1997}
}

@article{jordan1994hme,
  author  = {Michael I. Jordan and Robert A. Jacobs},
  title   = {Hierarchical mixtures of experts and the {EM} algorithm},
  journal = {Neural Computation},
  volume  = {6},
  number  = {2},
  pages   = {181--214},
  year    = {1994},
  doi     = {10.1162/neco.1994.6.2.181}
}

@article{li2023groups,
  author  = {Yasheng Li and Ningfang Liao and Yumei Li and Hongsong Li and Wenmin Wu},
  title   = {Color Conversion of Wide-Color-Gamut Cameras Using Optimal Training Groups},
  journal = {Sensors},
  volume  = {23},
  number  = {16},
  pages   = {7186},
  year    = {2023},
  doi     = {10.3390/s23167186}
}

@misc{middleburycolor,
  author       = {{Middlebury Computer Vision}},
  title        = {Middlebury Registered Color Checker Dataset},
  howpublished = {\url{https://vision.middlebury.edu/color/data/}},
  note         = {Dataset webpage, accessed April 4, 2026},
  year         = {2011}
}

\end{document}